\documentclass[12pt]{article}
\usepackage{amsmath,amssymb,amsthm,amsxtra,overpic,bbm,bm,epsfig,subfigure}
\usepackage[colorlinks=true,citecolor=blue,linkcolor=blue]{hyperref}
\usepackage{mathrsfs}
\usepackage{enumitem}
\usepackage{graphicx}
\usepackage{color}
\usepackage{comment}
\usepackage{epstopdf}
\usepackage{float}
\usepackage{cite}
\usepackage{multirow}
\usepackage{threeparttable}
\usepackage{slashed,stmaryrd}
\usepackage{lineno}

\def\thefootnote{\fnsymbol{footnote}}

\def\tabnotefont{\fontsize{9}{10}\selectfont}%

\begin{document}

\vspace{0.2cm}

\begin{center}
{\Large\bf A practical approach of high precision U and Th concentration measurement in acrylic}
\end{center}

\begin{center}
{\bf Chuanya Cao$^{a,b}$},
\quad
{\bf Nan Li$^{a,b}$}\footnote{Now at Huawei Technologies Co., Ltd.},
\quad
{\bf Xiaoyu Yang$^{a,c}$}\footnote{Email: yangxy@ihep.ac.cn},
\quad
{\bf Jie Zhao$^{a,c}$}\footnote{Email: zhaojie@ihep.ac.cn},
\quad
{\bf Yuanxia Li$^{a}$},
\quad
{\bf Zhiyan Cai$^{a,b}$},
\quad
{\bf Liangjian Wen$^{a,c}$}\footnote{Email: wenlj@ihep.ac.cn},
\quad
{\bf Xiaolan Luo$^{a,c}$},
\quad
{\bf Yuekun Heng$^{a,c}$},
\quad
{\bf Yayun Ding$^{a}$}
\\
\vspace{0.2cm}
{\small
$^a$Institute of High Energy Physics, Chinese Academy of Sciences, Beijing 100049, China\\
$^b$University of Chinese Academy of Sciences, Beijing 100049, China\\
$^c$State Key Laboratory of Particle Detection and Electronics, Beijing 100049, China\\}
\end{center}

\vspace{1.5cm}

\begin{abstract}
The Jiangmen Underground Neutrino Observatory will build the world's largest liquid scintillator detector to study neutrinos from various sources. The 20 kt liquid scintillator will be stored in a $\sim$600 t acrylic sphere with 35.4 m diameter due to the good light transparency, chemical compatibility and low radioactivity of acrylic~\cite{sec1:JUNOcdr}. The concentration of U/Th in acrylic is required to be less than 1 ppt (10$^{-12}$ g/g) to achieve a low radioactive background in the fiducial volume of the JUNO detector. The mass production of acrylic has started, and the quality control requires a fast and reliable radioassay on U/Th in acrylic. We have developed a practical method of measuring U/Th in acrylic to sub-ppt level using the Inductively Coupled Plasma Mass Spectrometer (ICP-MS). The U/Th in acrylic can be concentrated by vaporizing acrylic in a class 100 environment, and the residue will be collected and sent to ICP-MS for measuring U/Th. All the other chemical operation is done in a class 100 clean room, and the ICP-MS measurement is done in a class 1000 clean room. The recovery efficiency is studied by adding the natural nonexistent nuclei $^{229}$Th and $^{233}$U as the tracers. The resulting method detection limit (MDL) with 99\% confidence can reach 0.02/0.06 pg $^{238}$U/$^{232}$Th /g acrylic with $\sim$75\% recovery efficiency. This equipment and method can not only be used for the quality control of JUNO acrylic, but also be further optimized for the radioassay on other materials with extremely low radioactivity, such as ultra-pure water and liquid scintillator.
\end{abstract}

\begin{flushleft}
\hspace{0.9cm} Keywords: acrylic, ultra-low radioactivity, ICP-MS, uranium, thorium
\end{flushleft}

\def\thefootnote{\arabic{footnote}}
\setcounter{footnote}{0}

\newpage

\section{Introduction}
\label{sec:1}
Natural radioactivity from the detector material is always strictly controlled in low-background experiments due to its widespread and MeV level decay energy. Acrylic is widely used in large underground neutrino experiment~\cite{sec1:JUNOyellow,sec1:DYB,sec1:DC,sec1:RENO,sec1:SNO,sec1:EXO-200} and dark mater experiments~\cite{sec1:DEAP,sec1:LZ,sec1:DarkSide}, because acrylic has good transparency, low radioactive levels and good chemical compatibility.

The Jiangmen Underground Neutrino Observatory (JUNO) will build the world's largest liquid scintillator (LS) detector with 20 kt target mass to detect neutrinos from various sources. Acrylic is chosen as the spherical container of LS with 35.4 m inner diameter and 0.12 m thickness. Since the acrylic will be in contact with the LS directly, the requirement on the concentration of U/Th in acrylic is less than 1 ppt to avoid large radioactive background in the LS. The mass production of $\sim$600 t acrylic has started in 2019, and the quality control requires a fast and reliable radioassay of U/Th in acrylic. Gamma Spectroscopy (GS), Mass Spectrometer (MS), Alpha Spectroscopy (AS) and Neutron Activation (NA) are the main methods for the measurement of the radioactive nuclei. The detection limit of normal GS, mainly the High Pure Germanium (HPGe) detector, ranges from tens ppt to several ppb level for measuring U/Th~\cite{sec1:HPGe}. In order to reach high sensitivity, the acrylic can be irradiated by neutron or vaporized, and then either the excited daughters or the concentrated samples can by measured by GS or MS to sub-ppt level. It has to be noticed that another technique using Laser Ablation coupled to ICP-MS is also very promising~\cite{sec1:LA-ICPMS}. Several experiments and institutes have measured the U/Th in acrylic to sub-ppt level, as shown in Table~\ref{tab:compare}. Even though the NA analysis also has high sensitivity on measuring U/Th, the time spent as well as the cost are large. In addition, the JUNO acrylic is produced by Donchamp~\cite{sec1:Donchamp} in China. As a result, it is necessary to build a setup for measuring U/Th in acrylic to sub-ppt level in China to meet JUNO's requirement. Taking into account the time spent and expense, we finally choose the Vaporized+MS method for the design of the setup.

\begin{table}[htb]
\begin{center}
 \caption{Several measured results of U/Th concentration in acrylic.}
\renewcommand\arraystretch{1.3}
    \begin{tabular}{c|c|c|c|c}
            \hline
        Experiment & Technology & Institute & Acrylic mass & Measured U/Th (ppt)  \\ \hline
        SNO & NA+GS & CRL+NRU~\cite{sec1:SNOdata1,sec1:SNOdata2}& 800 g & $<$0.1   \\
        SNO & Vaporized+MS & CRL~\cite{sec1:SNOdata1,sec1:SNOdata2} & 6 kg & 0.2 / 0.5 \\
        DEAP-3600 & Vaporized+MS & CRL+NBS~\cite{sec1:DEAPdata} & 25 kg & $\sim$0.5   \\
        -- & NA+GS & INFN Milano-Bicocca~\cite{sec1:NAA} & 8.2 g & $<$0.3 / $<$0.5 \\
        -- & LA-ICPMS & CENBG+IPREM~\cite{sec1:LA-ICPMS} & 10 $\mu$g & $<$1 \\
          \hline
    \end{tabular}
	\label{tab:compare}
\end{center}
\end{table}

In this work, we use the furnace for vaporizing the acrylic in a class 100 environment, and the residue is collected and sent to ICP-MS for measuring U/Th. All the other chemical operation is done in a class 100 clean room, and the ICP-MS measurement is done in a class 1000 clean room. In addition, the calibration of the recovery efficiencies of U/Th during the pre-treatment is realized by adding the natural nonexistent $^{229}$Th and $^{233}$U as the tracers.

The rest of this paper is organized as follows: Section~\ref{sec:2} details the setup and method for vaporizing acrylic, the blank test and recovery efficiency analysis. Section~\ref{sec:3} shows some results and discussions on measuring U/Th in the acrylic panel of JUNO. Section~\ref{sec:4} is the summary of this study.

\section{Experimental section}
\label{sec:2}

%
%
\subsection{Pre-treatment process}
\label{sec:2.1}
The sample sent to ICP-MS should be liquid, and acrylic can be dissolved into concentrated nitric acid at high temperature and pressure using microwave digestion. However, the background level for this pre-treatment can reach 9$\sim$18 ppt due to the large contamination from the digestion vessel.  Another natural way for the pre-treatment is to vaporize the acrylic and leave behind the inorganic constituents, and the residue is easy to be dissolved in acid. The final design for the pre-treatment flow is shown in Figure~\ref{fig:flowchart} and briefly listed as follows:

\begin{figure}[!h]
\begin{center}
	\includegraphics[width=8cm]{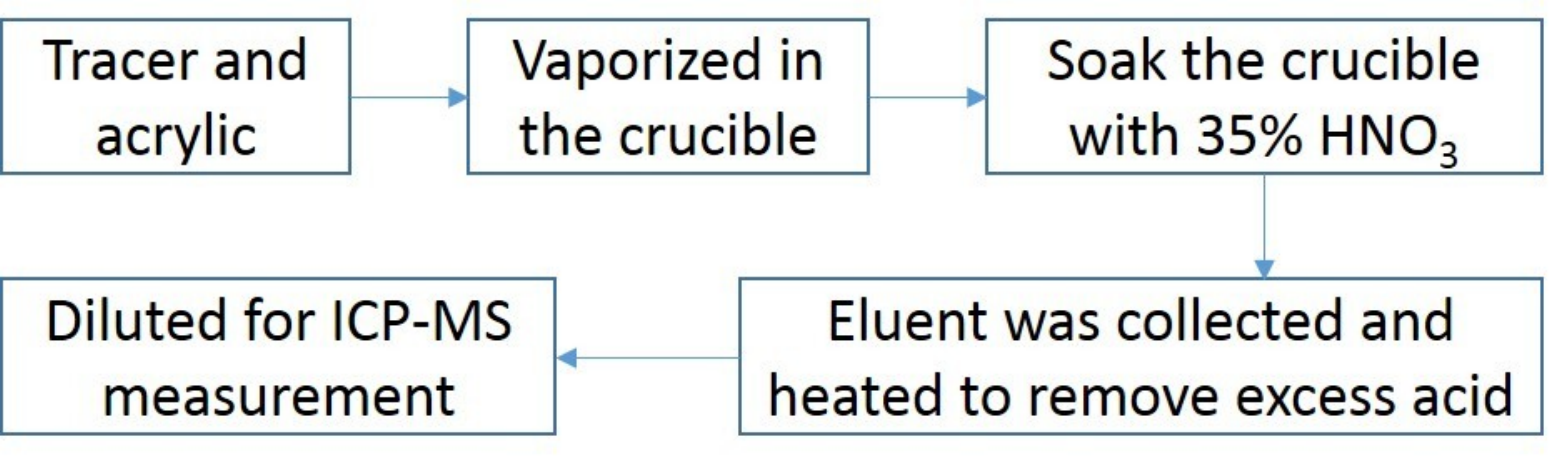}
	\caption{Flowchart of the pre-treatment process.}\label{fig:flowchart}
\end{center}
\end{figure}

\begin{itemize}
\item The natural nonexistent $^{229}$Th and $^{233}$U are taken as the tracers to obtain the recovery efficiency for the pre-treatment of acrylic. The acrylic sample as well as the tracers are put into the quartz crucible to avoid direct contact with the furnace tube.
\item Two furnaces are designed for the pretreatment of acrylic. The first one is for the vaporization and leaving behind the inorganic residue, the other is for the incineration of the organic components from the first furnace to avoid toxic gas pollution. The crucible with sample is put at the middle of the quartz tube of the first furnace.
\item The crucible is taken back to the ICP-MS lab after vaporization. The residue is collected by soaking the crucible with $\sim$35\% HNO$_3$ in the class 100 clean room.
\item The eluent was collected and heated to remove excess acid after acid soaking. Then the solution is diluted and sent to ICP-MS for the measurement.
\end{itemize}

Details about this pre-treatment will be discussed in the following sections.

\subsubsection{Facilities for the vaporization}
\par
The acrylic can be burned at high temperature, most light nuclei (C/H/O) can be driven away along with the gas, leaving behind the heavy nuclei (U/Th), which can be dissolved in nitric acid. The melting and burning point of acrylic are 130$^\circ$C and 458$^\circ$C. We designed a vaporization apparatus consisting of two burning furnaces, and the conceptual design is shown in Figure~\ref{fig:S3structure}. Detailed descriptions about this setup are listed below:

\begin{figure}[!h]
\begin{center}
	\includegraphics[width=0.9\textwidth]{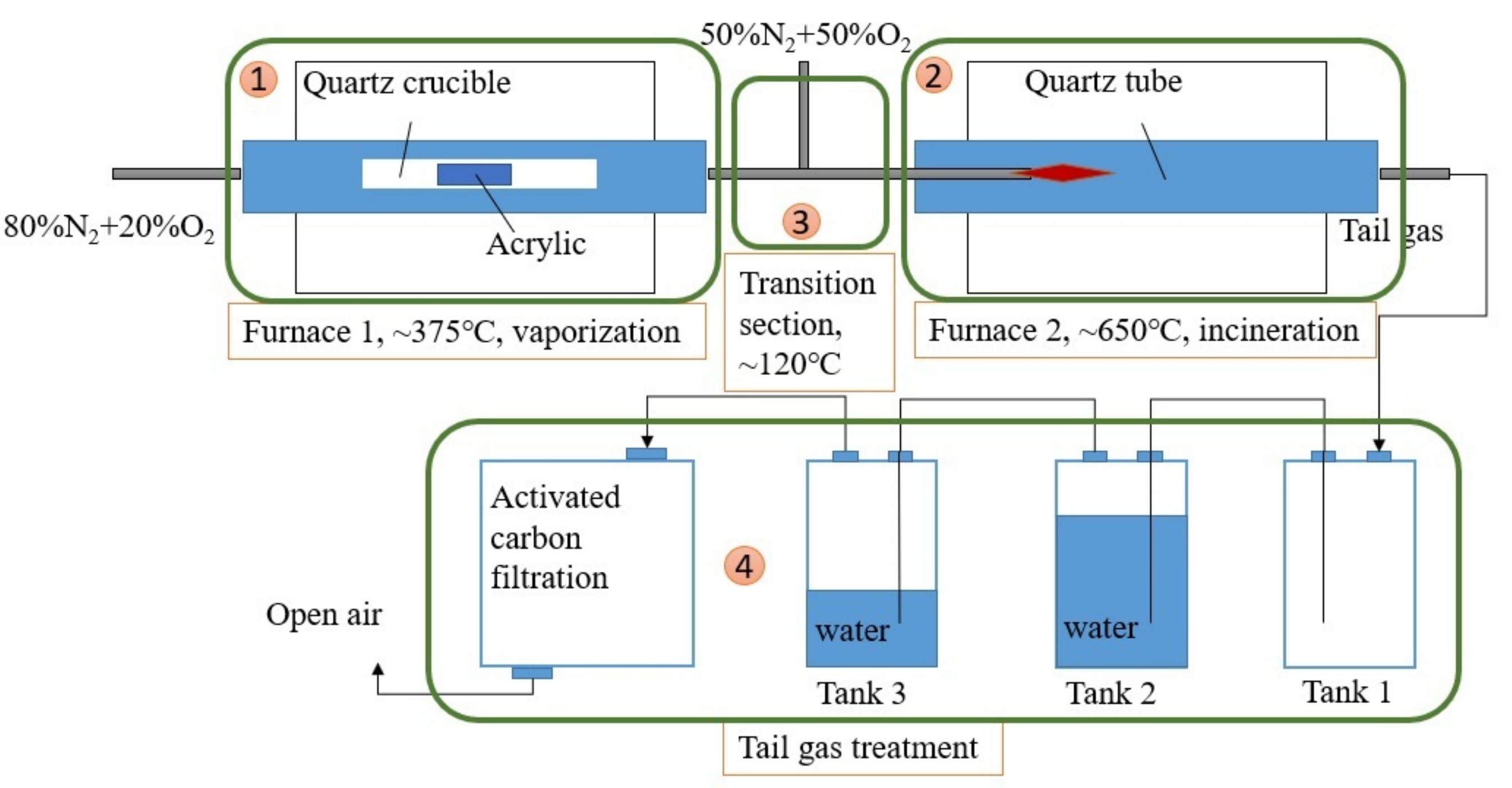}
	\caption{The conceptual design of the vaporization apparatus.}\label{fig:S3structure}
\end{center}
\end{figure}

\begin{itemize}
\item The first furnace (part 1 in Figure~\ref{fig:S3structure}) is used for acrylic boiling, which will remove the organic component and leave behind most of the inorganic residues. The acrylic sample is placed in a quartz crucible to avoid direct contact with the furnace tube. The gas is changed from N$_2$ only to 80\% N$_2+$ 20\% O$_2$ to avoid carbon accumulation on the quartz tube. The gas flow rate is tuned from 0.5 L/min to 2 L/min, and the optimized value is 1 L/min with no visible carbon accumulation. The working temperature is tuned between 340$^\circ$C and 375$^\circ$C, and the time spent at different temperature are shown in Table~\ref{tab:temFirstStage}. The lower temperature will lead to longer time spent. In addition, if this temperature is set to be higher than 400$^\circ$, the monomer can be fired in this furnace. The final value is set to 375$^\circ$C considering the time spent and avoid firing in this furnace.

\begin{table}[!h]
\begin{center}
	\caption{The time spent for acrylic boiling in the first furnace at different working temperature. The time for increasing the temperature from room level to the working condition can be controlled by setting the speed and is not included in this table.}
\renewcommand\arraystretch{1.3}
	\begin{tabular}{c|cccc}
        \hline
         Tempreture($^\circ$C)& 340 & 350 & 360 & 375 \\ \hline
         Time(min) & $>150$ & 120 & 80 & 60 \\
         \hline
	\end{tabular}
	\label{tab:temFirstStage}
\end{center}
\end{table}

    Even through most of acrylic is vaporized and carried off by the inlet gas, there exist little organic residue on the surface of the crucible after one hour of heating in the first furnace at 375$^\circ$C. In order to thoroughly remove the organic part, we have tried to increase the temperature at 600$^\circ$C for another half hour. No visible residue exists in the crucible after this operation, and the limited concentration of organic gas will not be fired in this furnace.

\item In order to save time, a triplet quartz crucible is designed to realize vaporizing three parallel samples at the same time, as shown in Figure~\ref{fig:S3acrylic}. The size of each single crucible is 70 mm $\times$ 40 mm $\times$ 34 mm, and the dimension of acrylic sample is 35 mm $\times$ 20 mm $\times$ 20 mm. After vaporization, this crucible will be taken back to the ICP-MS lab and soaked with 35\% nitric acid in the class 100 clean room for at least 15 hours.

\begin{figure}[!h]
\begin{center}
	\includegraphics[width=6cm]{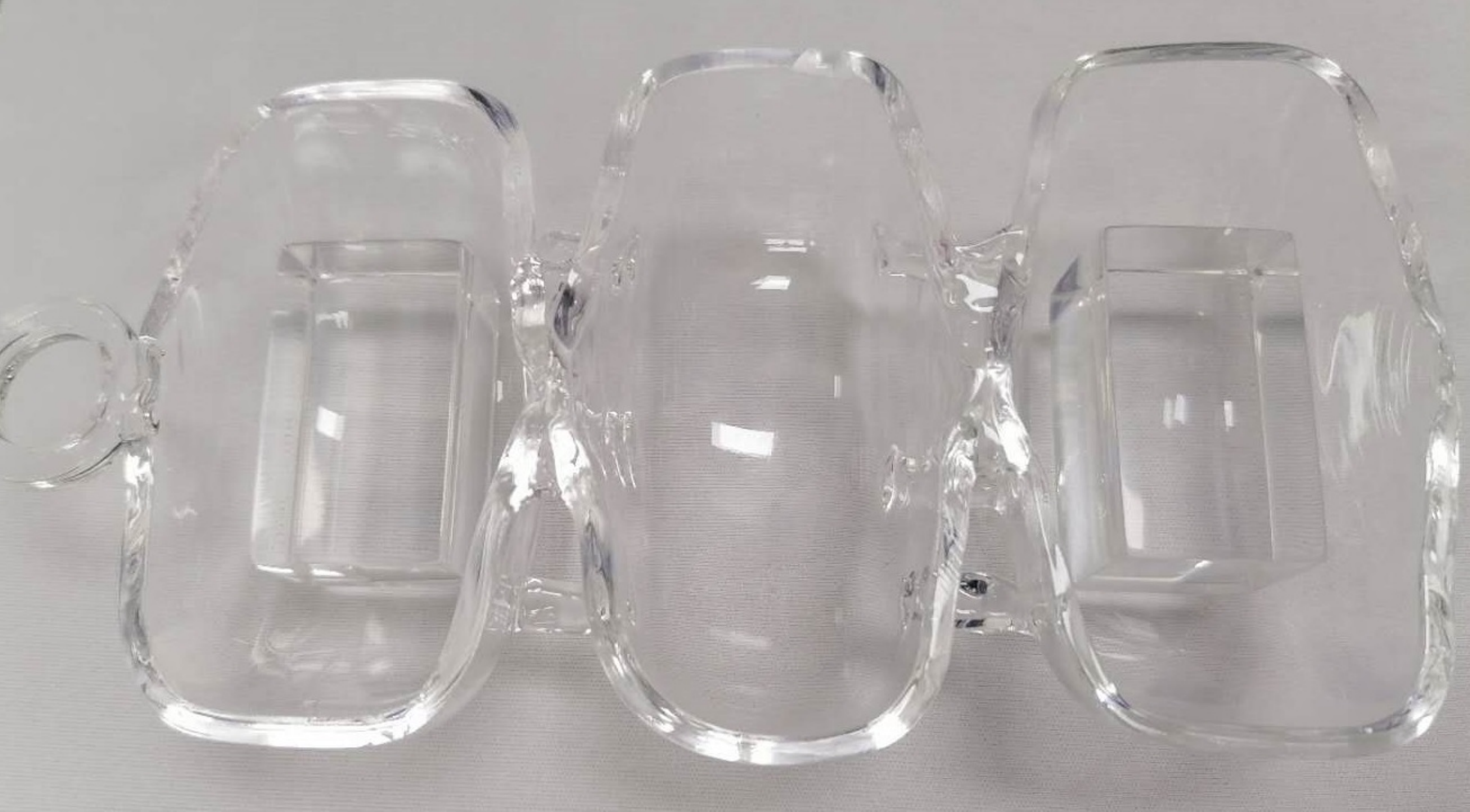}
	\caption{A triplet quartz crucible is designed to realize vaporizing three parallel samples at the same time. The dimension of each single crucible is 70 mm $\times$ 40 mm $\times$ 34 mm, while the dimension of acrylic sample is 35 mm $\times$ 20 mm $\times$ 20 mm.}\label{fig:S3acrylic}
\end{center}
\end{figure}

\item Since the acrylic monomer is poisonous and can not be discharged directly to the outside, the second furnace (part 2 in Figure~\ref{fig:S3structure}) is designed to burn the monomer with suitable amount of flowing oxygen, and convert it to harmless gas (CO$_2$ and H$_2$O) which can later be safely discharged to the open air. The casing cap should be used to keep stable and uniform temperature in the quartz tube (cylinder with 80 mm diameter and 750 mm length), however, this will lead to accumulation of acrylic monomer around the inlet. Explode can happen when these acrylic monomer is fired. In order to avoid this hidden danger, instead of using the casing cap, we extend the pipe of the transition section to the inner part of the second furnace, as shown in Figure~\ref{fig:S3furnace2}. By this way, the acrylic monomer with dense flow can be stably fired at 600$^\circ$C. The gas flow rate is tested from 2 L/min to 4 L/min at this furnace, and the optimized flow rate is 2 L/min with a stable incineration.

\begin{figure}[!h]
\begin{center}
	\includegraphics[width=6cm]{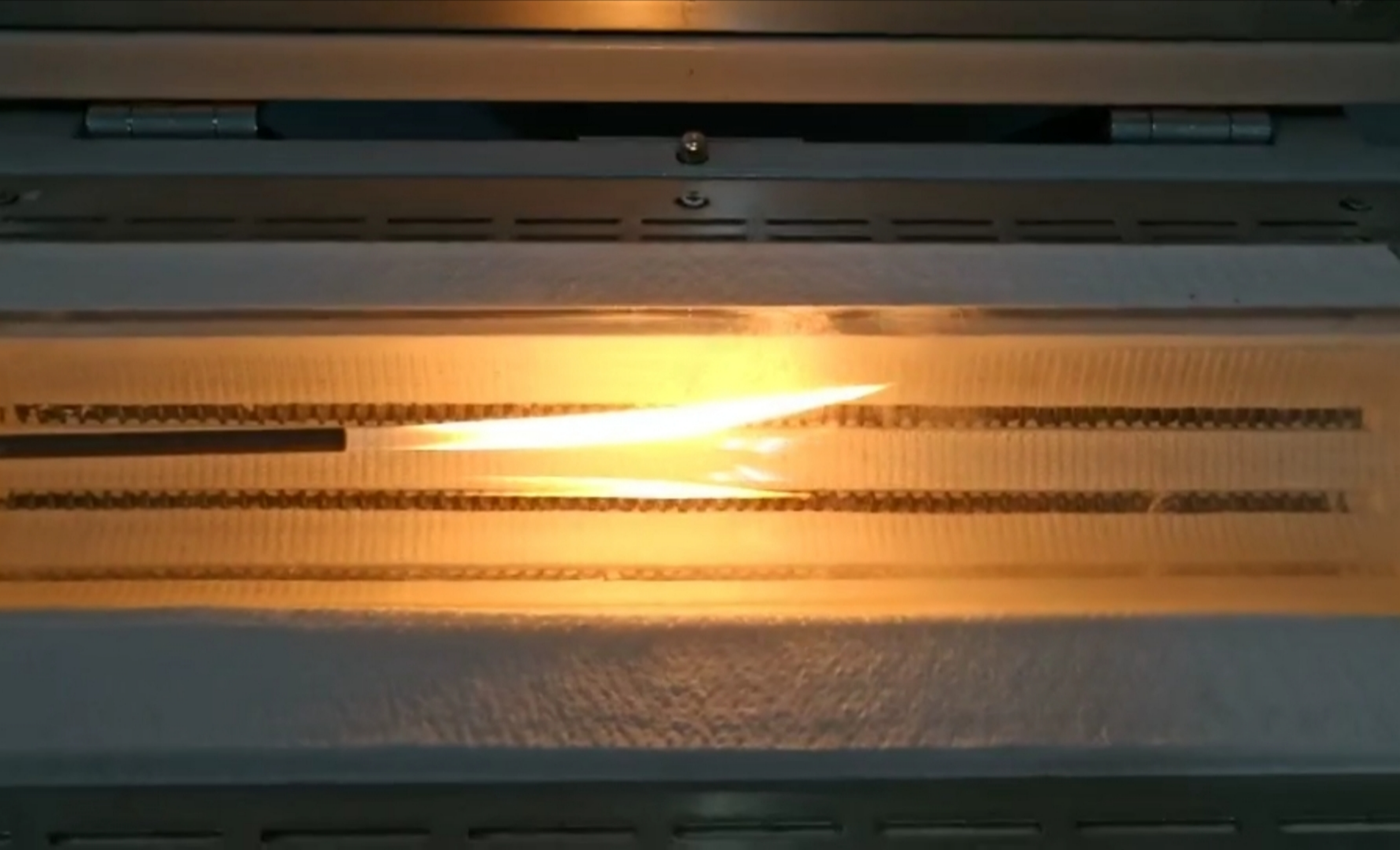}
	\caption{The design of the extend pipe from the transition section to the inner part of the quartz tube to achieve a stable firing with higher temperature and dense flow.}\label{fig:S3furnace2}
\end{center}
\end{figure}

\item The temperature of transfer section between the two furnaces (part 3 in Figure~\ref{fig:S3structure}) are kept at 120$^\circ$C to avoid condensation of acrylic monomer from the first furnace. The pipe size is enlarged from 4 mm to 12 mm to avoid block in the transition section.

\item In order to handle possible incomplete combustion of the monomer gas, special treatment on the tail gas is needed, which can also be seen in part 4 in Figure~\ref{fig:S3structure}. Tank 1 is used for overflow, while tank 2 and tank 3 are filled with water for gas cooling, then the residual gas will pass through the active carbon filtration.

\end{itemize}

\begin{figure}[!h]
\begin{center}
	\includegraphics[width=9cm]{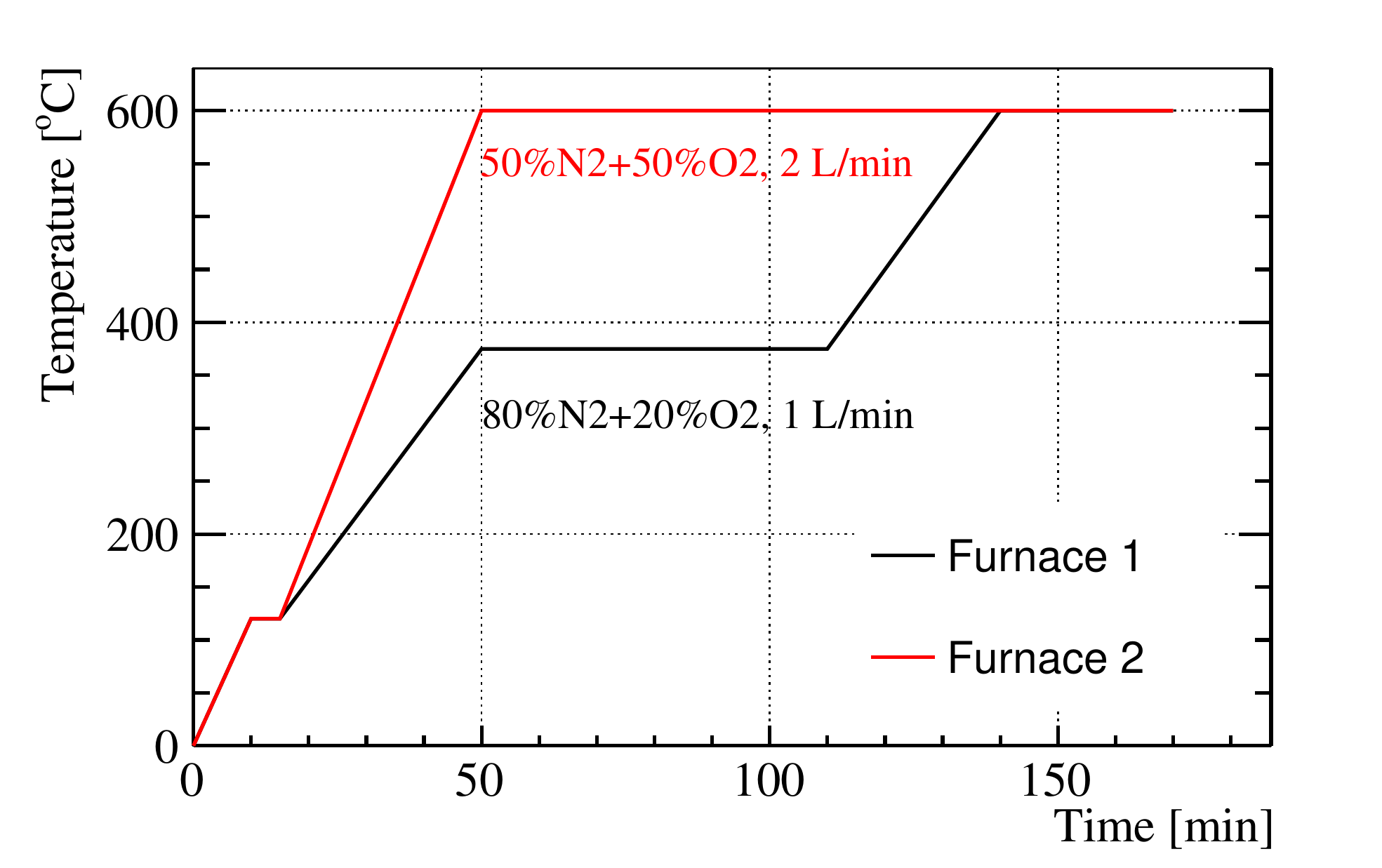}
	\caption{The final optimized operation parameters for vaporizing 15$\sim$50 g acrylic using furnaces. The furnace should be preheated at 120$^\circ$C for five minutes, then the temperature can rise to the working condition.}\label{fig:S3monitor}
\end{center}
\end{figure}

For 15$\sim$50 g acrylic, the final optimized operation parameters are shown in Figure~\ref{fig:S3monitor}. Different parameters may be tuned for other quantities of the acrylic to achieve thoroughly decomposition.

\subsection{Contamination control}
\label{sec:2.2}
\par
Careful control on the external contamination is essential in this work. The contamination mainly comes from the environment, container and reagent.

Since the U/Th in dust can reach ppm level, the clean operational environment is always needed. To decrease the effect of background collected during the vaporization, a clean tent is built and cleanliness level of 100 is reached with nobody in the tent. The cleanliness will be worsened by one order when there are human activities inside the tent. All the other chemical operations are done in a class 100 clean room, while the measurement is done in a class 1000 clean room. Besides that, the gas used for the vaporization is the mixture of N$_2$ and O$_2$ with 99.999\% purity.

\begin{table}[!h]
\begin{center}
\caption{The reagents used in this work.}
\begin{threeparttable}
	\begin{tabular}{cc}
        \hline
         Reagents or Labware & Description  \\ \hline
         ultra-pure wter & Milli-Q Reference \\
         detergent & Alconox \\
         nitric acid & OPTIMA (Fisher Scientific) \\
         $^{229}$Th standard\tnote{a} & 12.7 $ng\cdot$mL$^{-1}$ \\
         $^{233}$U standard\tnote{a} & 2.8 $\mu g\cdot$mL$^{-1}$ \\
         $^{238}$U/$^{232}$Th standard\tnote{b} & 100 $\mu g\cdot$mL$^{-1}$ \\
         \hline
	\end{tabular}
          \begin{tablenotes}
        \footnotesize
        \item[a] From Institute for Reference Materials and Measurements (IRMM).
        \item[b] From national standard reference material of P.R. China.
      \end{tablenotes}
    \end{threeparttable}
	\label{tab:reagents}
\end{center}
\end{table}

In order to reduce the pollution from the PFA vessels and quartz crucible, we have made a mature procedure for the cleaning of the containers. The reagents used in this work are listed in Table~\ref{tab:reagents}. Ultra-pure deionized water  (18~M$\Omega\cdot$cm) produced by the Milli-Q is used for cleaning. All the containers should be soaked in the detergent (Alconox) aqueous solution for at least 5 minutes under ultrasonic conditions, then rinsed with fresh water. In order to further remove the U/Th on the surface, the containers will then be soaked in the two-stage acid cylinders filled with 6 mol/L HNO$_3$ for at least one day at each stage. In the end, the containers will be filled with 6 mol/L HNO$_3$ and boiled for 10 minutes, then they are rinsed with fresh water and ready for use. All the standard solutions listed in Table~\ref{tab:reagents} are diluted with 2\% acid to ppt level for the ICP-MS calibration and recovery efficiency study.

After vaporization, the residue on the crucible is collected by soaking with 10 g $\sim$35\% nitric acid. Eluent is collected and heated to remove excess acid after at least 15 hours soaking, and the solution is diluted for ICP-MS measurement.

We have done various measurements on these sources and the results are summarized in Table~\ref{tab:contamination}. The required amount of reagents to deal with $\sim$16 g acrylic are also listed, and the fraction of contribution from each source is calculated in the last column. For other acrylic mass, the needed amount of reagent will be proportional to the mass. In Table~\ref{tab:contamination}, environmental contamination as well as the surface impurities from labware dominate the contamination. The contamination from labware can be further decreased by choosing sample container with lower radioactivity and better thermostability, such as platinum crucible.

\begin{table}[!h]
\begin{center}
\renewcommand\arraystretch{1.3}
\caption{Analysis of the external contaminations. For each reagent and labware, measured U/Th contamination and required amount to deal with $\sim$16 g acrylic are listed. Fractions represent relative contributions of each source in this assay. The total contaminated concentration is calculated on the bottom.}
	\begin{tabular}{c|cc|c}
        \hline
         Contamination Sources & $^{238}$U(pg) & $^{232}$Th(pg) & Fractions for U/Th \\ \hline
         HNO$_{3}$ (35\%, 10 g) & 0.03$\pm$0.02 & 0.06$\pm$0.01 & 12\% / 25\% \\
         2.4 ppt $^{229}$Th (1.5 g) & 0.05$\pm$0.01 & 0.05$\pm$0.01 & 19\% / 21\%  \\
         2.4 ppt $^{233}$U (1.5 g) & $<$0.01 & $<$0.01 & 4\% / 4\% \\
         Environment and labware & 0.17$\pm$0.06 & 0.12$\pm$0.03 & 65\% / 50\%  \\ \hline
         Total & 0.26 & 0.24 \\ \hline
	\end{tabular}
	\label{tab:contamination}
\end{center}
\end{table}

\subsection{Blanks $\&$ Method Detection Limit }
\label{sec:2.3}

\begin{figure}[!h]
\begin{center}
	\includegraphics[width=8cm]{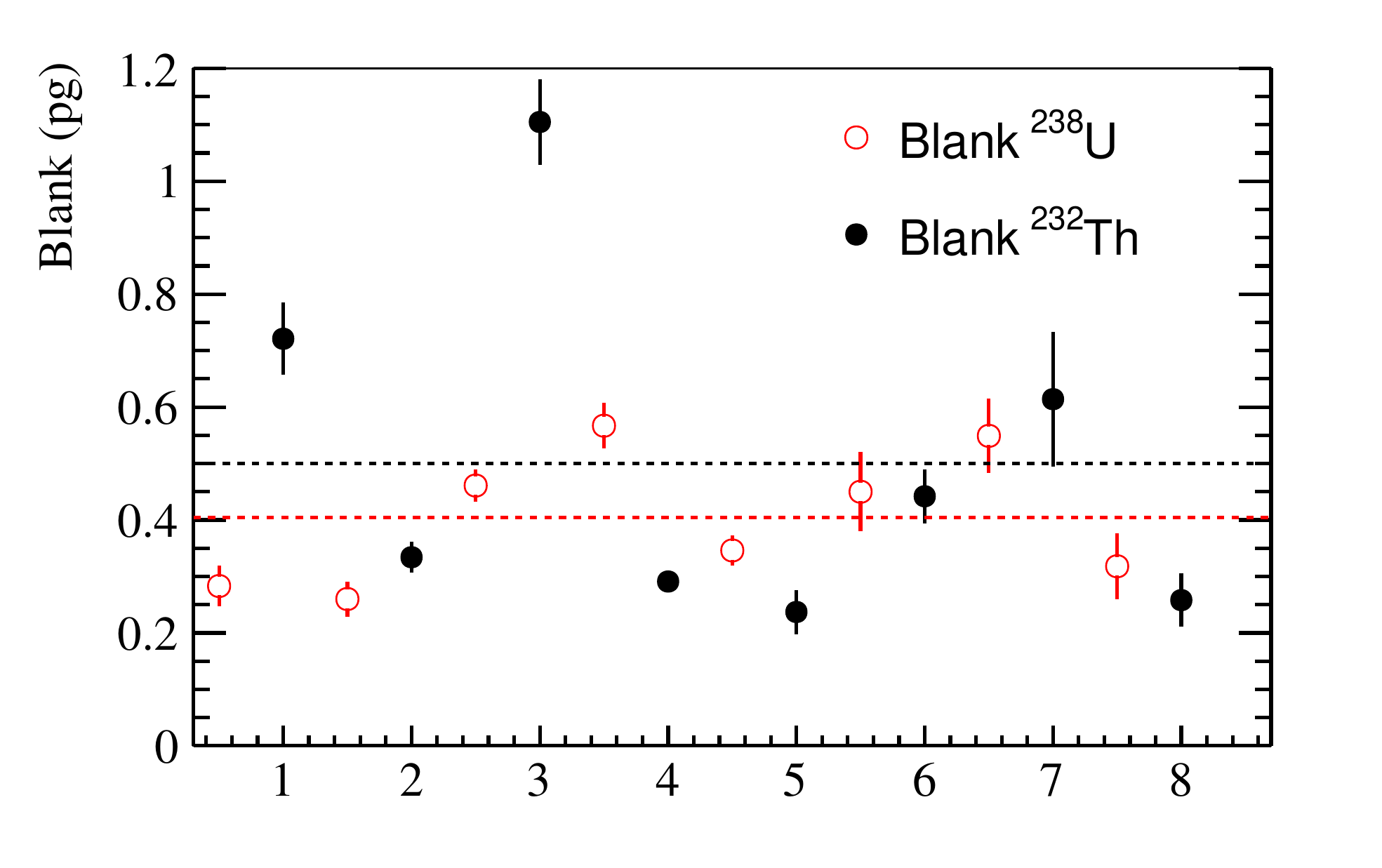}
	\caption{Eight parallel blanks have been prepared, and the measured U/Th concentration with recovery efficiency corrected are shown in this figure. The error bar for each point is the statistical uncertainty from ICP-MS measurement, and the average values are shown by dashed line.}\label{fig:S3blank}
\end{center}
\end{figure}

Instrumental analysis is important for the determination of the method detection limit (MDL) when any pre-treatment is involved. The conventional way is to prepare the blank solution, which should follow the same procedure as that of the sample, but without the sample. In this work, the blanks are prepared following the whole procedure of acrylic in Section~\ref{sec:2.1} and~\ref{sec:2.2} with all the reagent listed in Table~\ref{tab:contamination}. Eight parallel blanks have been prepared, and the measured U/Th concentration are shown in Figure~\ref{fig:S3blank}. The average value can reach (0.4$\pm$0.1) pg for $^{238}$U and (0.5$\pm$0.3) pg for $^{232}$Th, and the uncertainty is got by the RMS of the data points. The MDL with 99\% confidence is then calculated based on the definition in~\cite{sec3:MDL} to be 0.02/0.06 pg $^{238}$U/$^{232}$Th /g acrylic.

\subsection{Recovery efficiency}
\label{sec:2.4}
There may be some loss of U/Th during the pre-treatment, not only the vaporization but also the chemical operation. In order to get an accurate measurement of the U/Th in acrylic, the natural nonexistent nuclei $^{229}$Th and $^{233}$U are added to the sample for the recovery efficiency evaluation.

The mass of the tracer added to the sample should be at the same order with the mass of U/Th in the sample, since the recovery efficiency may be not stable at different concentration. About 16 g acrylic and 3.6 pg $^{229}$Th and $^{233}$U ($\sim$0.2 ppt) are put into the crucible, then gone through vaporization and sent to ICP-MS for measurement. The recovery efficiency can be calculated by comparing the measured mass of $^{229}$Th/$^{233}$U with the initial input.

Most organic components are removed by the vaporization, and U/Th in acrylic is less than 1 ppt, thus the total dissolved salts (TDS) in acrylic-samples were much less than 0.1\% by mass. As a result, external standard method is used for the calibration of ICP-MS. A series of $^{229}$Th, $^{232}$Th, $^{233}$U, $^{238}$U standards with different concentrations are prepared in advance. A linear function is used to fit the measured signals (in counts-per-second unit) versus different standard concentrations.

\begin{figure}[!h]
\begin{center}
	\includegraphics[width=8cm]{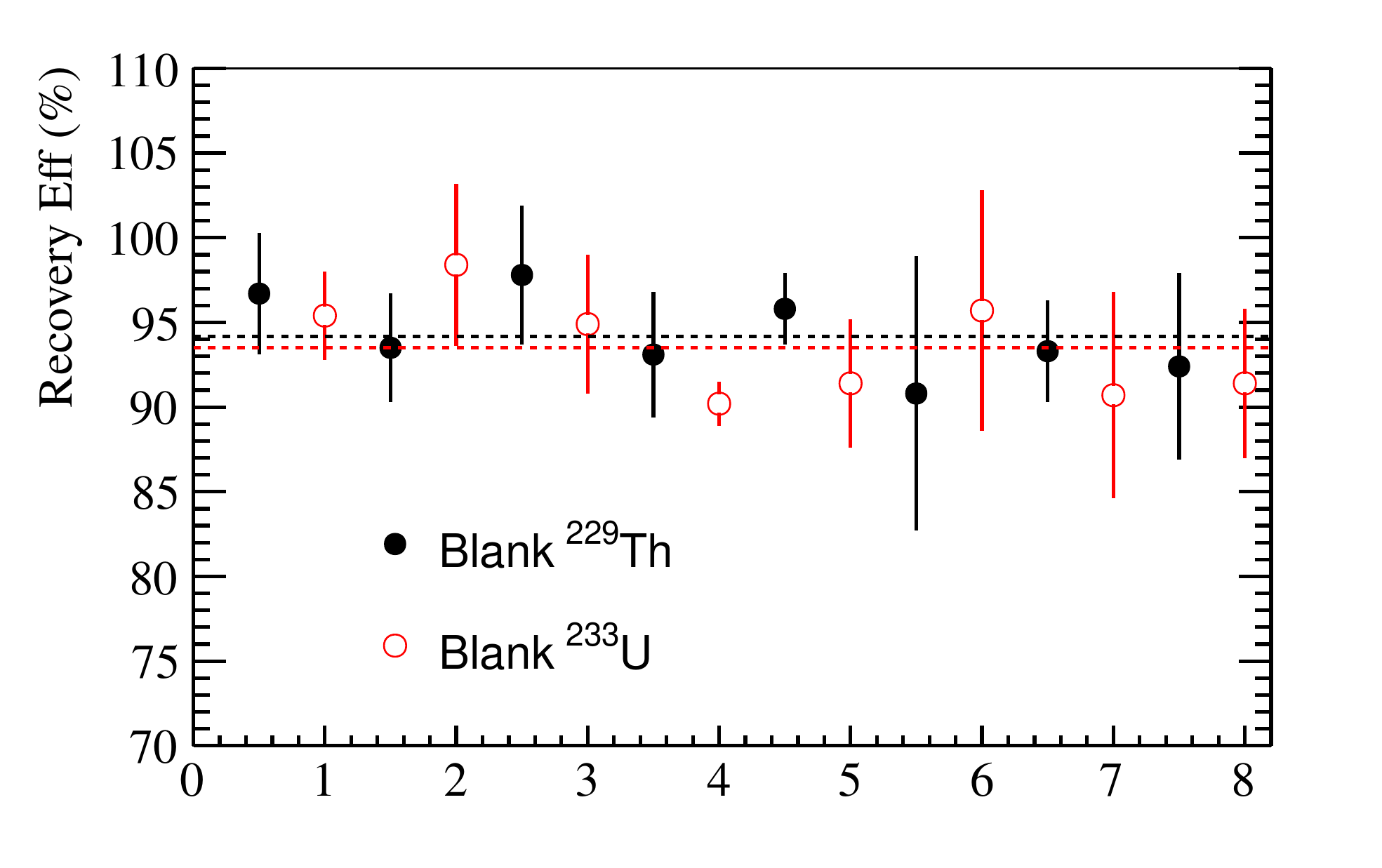}
    \includegraphics[width=8cm]{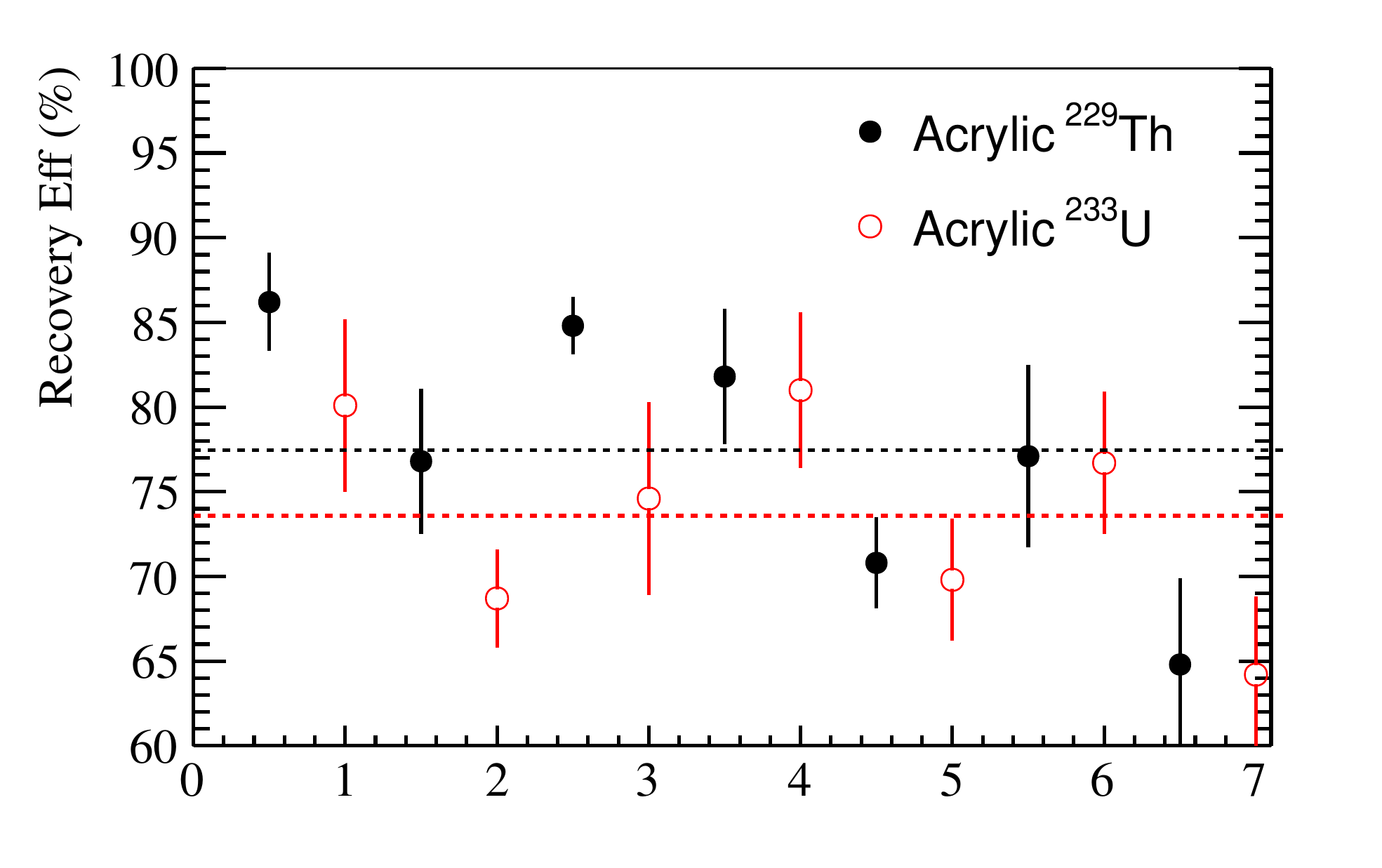}
	\caption{The recovery efficiencies of eight parallel blanks (left figure) and seven acrylic samples (right figure). The error bar for each point is the statistical uncertainty from ICP-MS measurement, and the average values are shown by dashed line.}\label{fig:S3eff}
\end{center}
\end{figure}

The recovery efficiencies of eight parallel blanks and seven acrylic samples are shown in Figure~\ref{fig:S3eff}, and only the statistical uncertainty from ICP-MS measurement is included in the data point. The average recovery efficiency for $^{229}$Th and $^{233}$U from the blank test are (94$\pm$3)\%, while they are (77$\pm$7)\% and (74$\pm$5)\% for acrylic, and the uncertainty is calculated using the RMS of the data points. Compared with the blank, the recovery efficiency of acrylic is lower. The reason is that we just use $\sim$10 g HNO$_3$ for the soaking to save time on removing excess acid, while the acrylic sample is $\sim$16 g. As a result, the residue on the crucible surface without contacting the acid will be lost. There is still some space for the optimization of sampling.

If the crucible is not clean enough, some U/Th in the surface can separate out during the vaporization. In order to estimate this effect, we have vaporized the acrylic without adding tracer, and measuring the $^{229}$Th/$^{233}$U in the solution. The results show no $^{229}$Th/$^{233}$U in the sample solution, that means our cleaning procedure is good enough for the crucible and vessels.

\section{Results and discussions}
\label{sec:3}
\par
\subsection{Acrylic results}
\label{sec:3.1}
Several acrylic samples from the acrylic panel in JUNO have been measured. The samples have been soaked in the detergent aqueous solution for 5 minutes under ultrasonic conditions, then rinsed with fresh water. After vaporization, the residue is collected using the method discussed in Section~\ref{sec:2.2} and sent to ICP-MS for measurement. The concentration of U/Th in acrylic can be calculated by

\begin{eqnarray}
C_i=(\frac{M_{i}*m_s}{\varepsilon}-B_{i})/m_a
\end{eqnarray}

where $M_{i}$ is the measured concentration of $i$th isotope; $B_{i}$ is the additional mass of $i$th isotope got from the blank test, as shown in Section~\ref{sec:2.3}; $\varepsilon$ is the recovery efficiency for the pre-treatment of acrylic, and will be measured in real time; $m_s$ is the mass of the sample solution sent to ICP-MS, and $m_a$ is the acrylic mass.

\begin{figure}[!h]
\begin{center}
	\includegraphics[width=8cm]{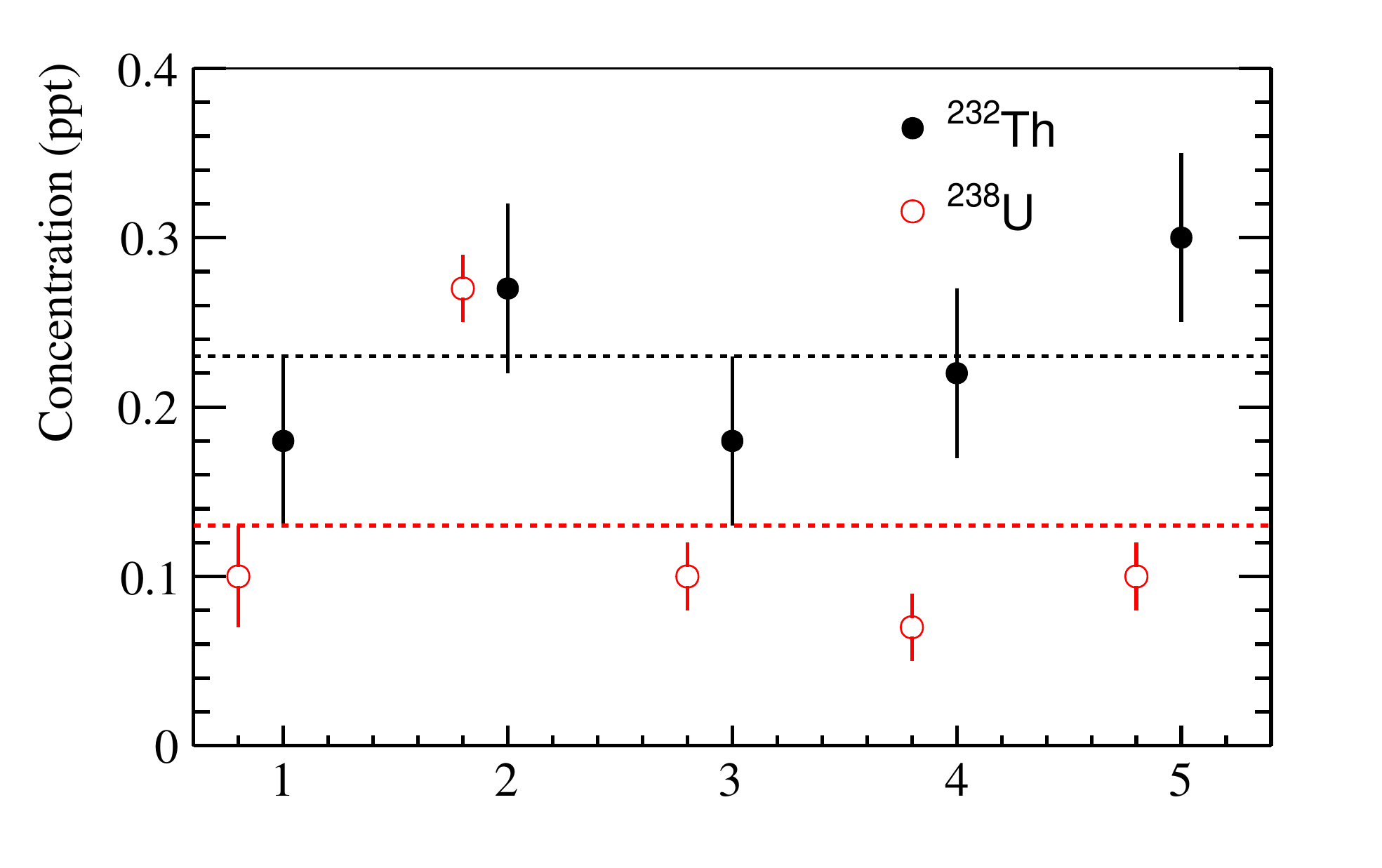}
    \caption{The measured U/Th concentration of five parallel samples from one acrylic panel in JUNO are shown as the full black circle for $^{232}$Th and empty gray circle for $^{238}$U in the figure. The data points are got with background and recovery efficiency corrected, and the uncertainty is got from the statistics of ICP-MS measurement. The average values are shown by dashed line.}\label{fig:S3junoAcrylic}
\end{center}
\end{figure}

We have got five parallel samples from one acrylic panel in JUNO with similar position and same cleaning procedures, and the results are shown in Figure~\ref{fig:S3junoAcrylic} with background and recovery efficiency corrected. The uncertainty of the data points are calculated from the statistics of ICP-MS measurement. The RMS of the five parallel results are 0.07 ppt for U/Th.

\subsection{Discussions}
\label{sec:3.2}
\par
The result in Section~\ref{sec:3.1} shows that the radioactivity of U/Th in JUNO acrylic can reach the requirement ($<$1 ppt). However, the acrylic can be easily contaminated by the environment during the complicated steps of production, such as bending. The flat acrylic panel is first produced, and the panel will be bent to the final product by heating and machining. Fortunately, the acrylic panel will be polished and cleaned in the end, and the key point is the procedure of the cleaning. The materials and reagent used for sanding, griding and polishing have been carefully screened by the HPGe detector. Several tests have been done on the first bent acrylic panel with different cleaning steps, and the results are shown in Table~\ref{tab:acrylicDataClean}.

\begin{table}[!h]
\begin{center}
\footnotesize
\renewcommand\arraystretch{1.3}
	\begin{tabular}{c|cc|cc}
        \hline
         \multirow{2}{*}{Unit:ppt} & \multirow{2}{*}{$^{238}$U} & \multirow{2}{*}{$^{232}$Th} & \multicolumn{2}{c}{Treatment} \\
         \cline{4-5}
         &&& Company & Lab \\ \hline
         No.1 & 2.41$\pm$0.16 & 5.91$\pm$0.29 & $S+P+AC$ & $WC$ \\
         No.2 & 0.44$\pm$0.05 & 1.01$\pm$0.05 & $S+P+AC$ & $AC$ \\ \hline
         No.3 & 5.12$\pm$0.09 & 7.28$\pm$0.11 & $S+G+P+AC$ & $WC$ \\
         No.4 & 0.69$\pm$0.05 & 3.09$\pm$0.18 & $S+G+P+AC$ & $AC$ \\ \hline
         No.5 & 0.19$\pm$0.03 & 0.57$\pm$0.03 & $S+G+P+AC$ & $WC$ \\
         No.6 & 0.33$\pm$0.02 & 1.12$\pm$0.07 & $S+AC+G+P+AC$ & $WC$ \\
         No.7 & 0.20$\pm$0.02 & 0.72$\pm$0.05 & $S+WC+G+P+AC$ & $WC$ \\
         \hline
	\end{tabular}
	\caption{The measured results of several tests on the first acrylic bending panel in JUNO with different operation process and cleaning steps, and they are sorted by sampling time. $S$, $G$, $P$ represent the sanding, griding and polishing process, while $AC$ and $WC$ represent Alconox solution Cleaning and ultra-pure Water Cleaning.}
	\label{tab:acrylicDataClean}
\end{center}
\end{table}

Some conclusions can be summarized as follows:

\begin{itemize}
\item From No.1(3) and No.2(4), there exist surface contamination that can be removed by Alconox solution. That means the radioactivity from the acrylic panel itself can reach the requirement, while there exists problems in the cleaning procedure at company.
\item From No.2 and No.4, it seems the operation of griding will add more contamination on the acrylic surface especially for Th.
\item From No.5, No.6 and No.7, after revising the cleaning procedure at company, all three samples can reach the requirement. Besides that, it seems less manual operation between sanding and polishing will lead to less contamination. We finally choose the procedure on the sample No.5 as our default operation process.
\end{itemize}

This technique can measure the U/Th in acrylic to sub-ppt level quickly and effectively, and it will play an important role on the quality control of acrylic mass production and operation process.

\section{Summary}
\label{sec:4}
In this study, we have successfully built the method for measuring U/Th in acrylic to sub-ppt level. The acrylic is vaporized in class 100 environment, and the residue is collected and measured by ICP-MS in a class 100 clean room. All the other chemical operations are done in the class 100 clean room to avoid dust contamination. The blank test shows that the MDL with 99\% confidence is 0.02/0.06 pg $^{238}$U/$^{232}$Th /g acrylic. The recovery efficiency is studied by adding the natural nonexistent nuclei $^{229}$Th and $^{233}$U, and the results show (74$\pm$5)\% and (77$\pm$7)\% recovery efficiency for the whole pre-treatment procedure. Several samples of the acrylic panel in JUNO have been vaporized and measured, and the results show the acrylic can reach JUNO's requirement ($<$1 ppt). Several quality control on the operation and cleaning procedure are done, which will give guidelines to the process optimization in the company.

There are still some space for further optimization. The blank level can be further decreased by choosing container with lower radioactivity and better thermostability, such as platinum crucible. The recovery efficiency of acrylic can be improved by optimizing the method of sampling. Besides that, this equipment can also be optimized for radioassay on other materials with extremely low radioactivity, such as ultra-pure water and liquid scintillator.

\section{Acknowledgement}
\par
This work is supported by the National Natural Science Foundation of China (Grant No. 11820101005), the Strategic Priority Research Program of the Chinese Academy of Sciences (Grant No. XDA10010200), and the National Natural Science Foundation of China (No. 11905226).

\end{document}